\documentclass{article}

\usepackage{authblk}
\usepackage{graphicx}
\usepackage{epstopdf}
\usepackage[pagewise]{lineno}
\usepackage{verbatim} 
\usepackage{amsmath}
\usepackage{hyperref}
\usepackage{url}
\usepackage{bm}
\usepackage[numbers]{natbib}
\usepackage{caption}
\usepackage{fancyhdr} 
\usepackage{url}

\fancypagestyle{plain}{%
  \fancyhf{} 
  \fancyfoot[R]{\thepage} 
}

\let\citeleft=(
\let\citeright=)

\begin{document}

\setcounter{section}{1} 

\pdfinfo{
   /Author (Suma Anand)
   /Title (Beat Pilot Tone (BPT): Simultaneous MR Imaging and RF Motion Sensing at Arbitrary Frequencies)
}

\title{\vspace{-2cm} Beat Pilot Tone (BPT): Simultaneous MR Imaging and RF Motion Sensing at Arbitrary Frequencies}

\author[1]{Suma Anand}
\author[2]{Michael Lustig}

\affil[1]{\small Electrical Engineering and Computer Sciences, University of California, Berkeley}
\affil[2]{\small Electrical Engineering and Computer Sciences, University of California, Berkeley}
\maketitle

\vfill
\noindent

\noindent
\textit{Address correspondence to:} \\
  506 Cory Hall, Berkeley, CA, 94720 \\
  mikilustig@berkeley.edu

\noindent
This work was supported by GE Healthcare, the NSF GRFP fellowship, and NIH U01EB025162, R01HL136965, and U01EB029427.

\noindent
Approximate word count: 246 (Abstract) 6661 (body)\\

\noindent
Submitted to \textit{Magnetic Resonance in Medicine} as a Research Article.

\clearpage


\section*{Abstract}
\noindent
\textbf{Purpose}: To introduce a simple system exploitation with the potential to turn MRI scanners into general-purpose RF motion monitoring systems.

\noindent
\textbf{Methods}: Inspired by Pilot Tone (PT), this work proposes Beat Pilot Tone (BPT), in which two or more RF tones at arbitrary frequencies are transmitted continuously during the scan. These tones create motion-modulated standing wave patterns that are sensed by the receiver coil array, incidentally mixed by intermodulation in the receiver chain, and digitized simultaneously with the MRI data. BPT can operate at almost any frequency as long as the intermodulation products lie within the bandwidth of the receivers. BPT’s mechanism is explained in electromagnetic simulations and validated experimentally.

\noindent
\textbf{Results}: Phantom and volunteer experiments over a range of transmit frequencies suggest that BPT may offer frequency-dependent sensitivity to motion.

\noindent Using a semi-flexible body receiver array, BPT appears to sense cardiac-induced body vibrations at microwave frequencies ($ \geq 1.2$ GHz). At lower frequencies, it exhibits a similar cardiac signal shape to PT, likely due to blood volume changes. 

Other volunteer experiments with respiratory, bulk, and head motion show that BPT can achieve greater sensitivity to motion than PT and greater separability between motion types. Basic multiple-input multiple-output ($4\times 22$ MIMO) operation with simultaneous PT and BPT in head motion is demonstrated using two transmit antennas and a 22-channel head-neck coil.

\noindent
\textbf{Conclusion}: BPT may offer a rich source of motion information that is frequency-dependent, simultaneous, and complementary to PT and the MRI exam.
  
\noindent
\textbf{Keywords}: radio frequency, motion sensing, microwave, pilot tone

\clearpage


\section*{Introduction}
\label{sec:introduction}
Motion is the most common unanticipated event in a clinical MRI examination \cite{sadigh2017prevalence}. Even when anticipated, such as with respiratory and cardiac motion, motion during the examination degrades image quality, resulting in repeated exams and increased costs \cite{Runge2019}. Moreover, as spatial resolution continues to improve, motion is accentuated and causes greater image corruption \cite{Zaitsev2015, Maclaren2013, Godenschweger2016}.

\begin{figure*}[!htb]%
\centering
\includegraphics[width=0.6\textwidth]{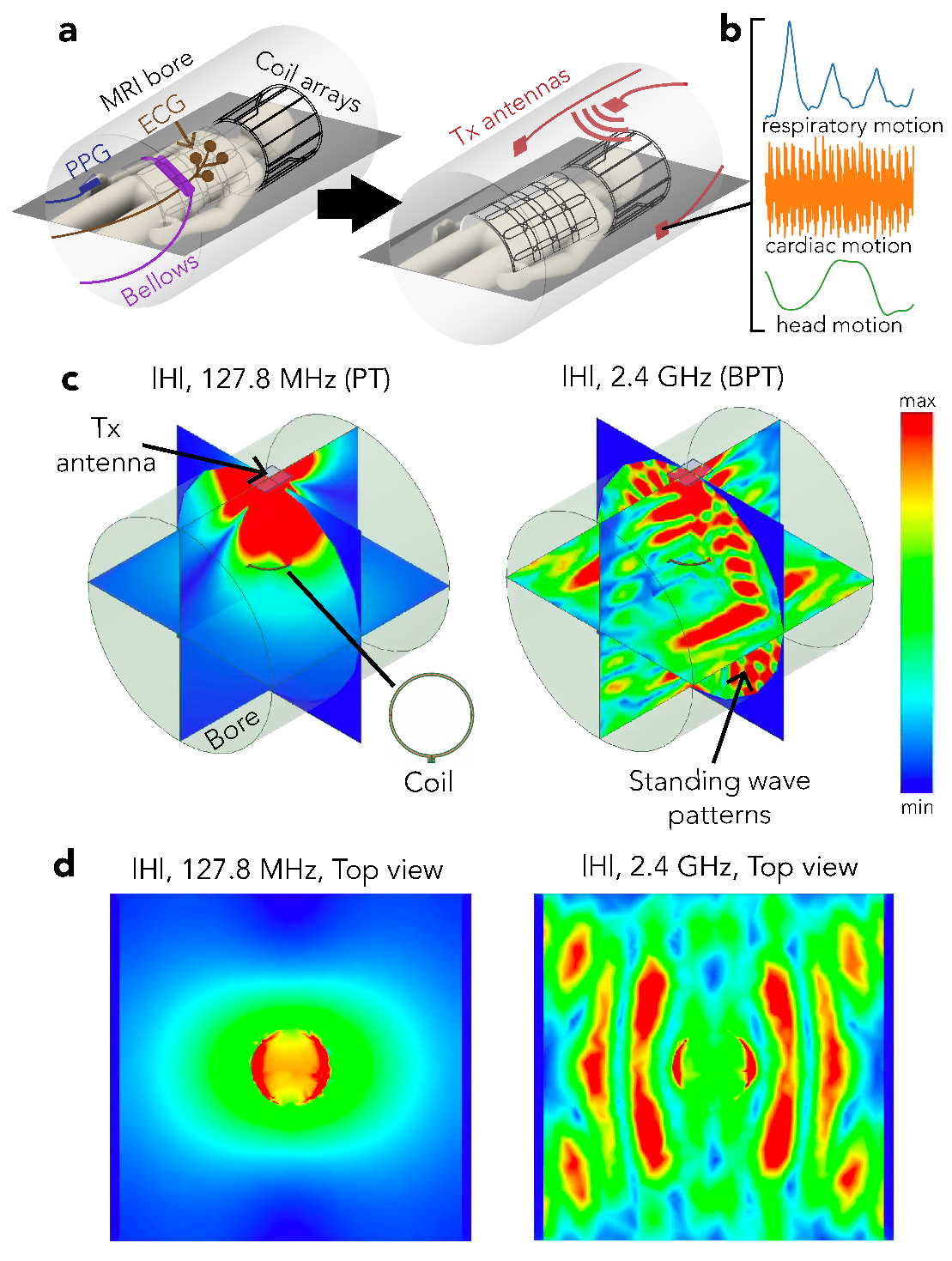}
\caption{\textbf{BPT concept}. a) Conventional MRI suite sensors (left) are bulky and require contact with the patient, while our method (right) utilizes MRI coil arrays and requires a transmit (tx) antenna in or near the bore to provide motion estimates (b). Multiple antennas can be used for MIMO operation. c) Simulated magnitudes of the H-field generated by a tx dipole antenna at $127.8$ MHz (left) and $2.4$ GHz (right), windowed to the same relative levels. A receiver coil tuned to $127.7$ MHz is placed $10$cm above the center of the bore. At $2.4$ GHz, standing wave patterns emerge. d) Top view of $|H|$ at (left) 127.8 MHz and (right) 2.4 GHz in the plane of the coil.
}
\label{fig:concept}
\end{figure*}

Simple motion mitigation techniques, like breath-holding, may prove unreliable and limit scan options. For instance, 3-dimensional imaging requires longer scan-time than a single breath-hold duration. Instead, modern motion correction techniques require measuring the motion during the exam and correcting artifacts prospectively, retrospectively, or both. While it is possible to use only the MR data itself for correction \cite{pipe1999motion, prieto2007reconstruction, Huttinga2020}, external motion monitoring signals or images offer a rich source of information that may improve the quality and speed of the correction \cite{Madore2022, huttinga2023gaussian, eschelbach2019comparison, stam2012navigators, Cheng2015, odille2008generalized, mcclelland2013respiratory, white2010promo}.

Existing motion sensing systems in MRI may have limitations in sensitivity, ease of use, or patient comfort. Cameras \cite{Maclaren2012,VanderKouwe2021} are accurate but require a line-of-sight path that is often blocked by the MRI coils and are limited in penetration depth \cite{Madore2022}. NMR and MRI-based navigators \cite{maclaren2013prospective, Godenschweger2016, white2010promo, stam2012navigators, Cheng2015, pipe1999motion} require sequence modifications and are adopted in limited applications \cite{frost2023k}. RF sensing \cite{buikman1988rf, roemer1990nmr, Andreychenko2017, Hess2018a, speier2015pt, Bacher2017, Thiel2009b, Wang2021, Li2013} can detect a variety of motion types with high sensitivity and without contact; however, existing solutions are lacking in either sensitivity or generality. ``RF coil" (RFC) sensors \cite{buikman1988rf, Andreychenko2017, Hess2018a, speier2015pt} use the  MRI receiver or transmitter coils as motion sensors but are tied to the Larmor frequency, thus limiting their inherent sensitivity, while ultra-high frequency (UHF) RF sensors such as radars \cite{Thiel2009b, Wang2021, Li2013} require significant engineering efforts to be integrated with the MRI scanner, hindering general implementation.

In this paper, we introduce Beat Pilot Tone (BPT), a simple system exploitation with the potential to turn any MRI scanner into a general-purpose RF motion monitoring system. BPT involves transmitting RF tones that interact with the body and MRI bore to create motion-modulated standing wave patterns (Figure \ref{fig:concept}c-d). These waves are sensed by the receiver coil array, incidentally mixed through nonlinear intermodulation in the receiver chain, and digitized simultaneously with the MRI data. BPT can therefore operate at almost any frequency, as long as the intermodulation lies within the bandwidth (BW) of the receivers. To distinguish between the transmitted and received BPT, we denote the entire concept as ``BPT", the transmit BPT field as ``BPT-Tx," and the sensed BPT signals after mixing as ``BPT-Rx". Figures \ref{fig:concept}c and d show a simulated BPT-Tx at 2.4 GHz for a transmitter placed at the top of the bore and a receiver coil 10 cm above the center.

Expanding the transmit frequency range offers the potential for frequency-dependent motion sensitivity; however, understanding its mechanism requires modeling electromagnetic (EM) wave effects at higher frequencies. We model these effects in simulation and validate them in experiment over a range of transmit frequencies ($127.8$ MHz --- $2.5278$ GHz). We show that the standing wave patterns from BPT-Tx increase in complexity and percent modulation with frequency. We evaluate the frequency-dependent motion sensitivity of BPT-Rx in phantom and volunteer experiments for common motion types (respiratory, bulk, cardiac, and head motion) using different coil arrays. BPT-Rx at microwave frequencies has the potential to separate motion types more easily than a well-established RFC method known as Pilot Tone (PT), which uses a single RF tone at $127.8$ MHz on our 3T MRI scanner. Moreover, BPT-Rx cardiac signal on volunteers reflects blood volume changes below 1.2 GHz and correlates strongly with small body vibrations (displacement ballistocardiogram; dBCG) at higher frequencies using a semi-rigid array. Finally, we show the potential for quantitative motion correction when using BPT as a frequency-multiplexed MIMO system. With two transmitters placed at different locations, BPT-Rx shows differences between two different head motions (nodding “yes” and shaking “no”) in a pattern similar to motion derived from image registration. This approach has been demonstrated for retrospective head motion correction in preliminary experiments \cite{Huttinga2023}. BPT thus offers the potential for simultaneous MR imaging and motion monitoring at arbitrary frequencies in many applications.


\section*{Methods}
\label{sec:methods}

We describe the general method of BPT (Section \ref{sec:bpt_rx}) and its hardware implementation (Section \ref{sec:hw}). We then hypothesize where and how BPT-Rx is created (Section \ref{sec:intermod}). We examine the mechanism of BPT in EM simulations (Section \ref{sec:sim}) and experiments (Section \ref{sec:sensitivity}) which show its sensitivity to small vibrations. 

\subsection{BPT Signal Reception}
\label{sec:bpt_rx}

\begin{figure}[!htb]
\centering
\includegraphics[width=0.45\textwidth]{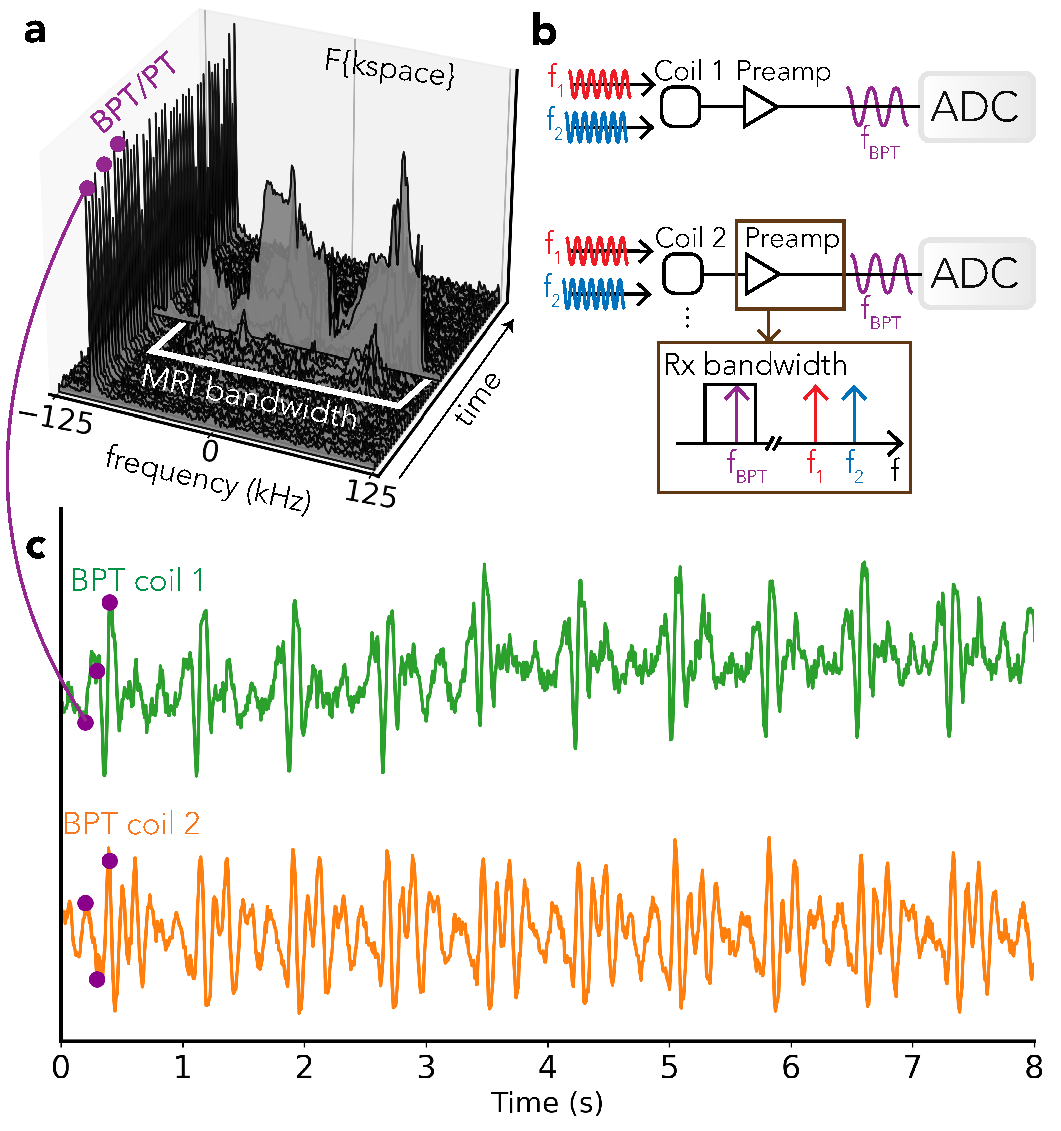}
\caption{\textbf{Pipeline for BPT extraction}. a) BPT-Rx appears as a peak in the Fourier transform of each \textit{k}-space line. The intermodulation frequency $f_{BPT}$ falls within the BW of the receiver but outside the BW of the MR image data. b) Model of the receiver chain: the electromagnetic fields at frequencies $f_1$ (red) and $f_2$ (blue) are sensed by MRI receiver coils and mixed by the preamplifiers to generate an intermodulation product at frequency $f_{BPT}$ \cite{Anand2021, Lamar2022}. c) Example of the magnitude of the BPT-Rx signal from two receiver coils, corresponding to the time-frequency plot in a) and showing amplitude modulation due to cardiac motion in a breath-holding volunteer.}
\label{fig:bpt_acq}
\end{figure}

BPT builds off of PT \cite{speier2015pt, Bacher2017}; however, BPT is additionally sensitive to electromagnetic wave effects, potentially allowing detection of subtler movements. RFC methods share a similar mechanism for sensing motion. Motion changes the load impedance to the transmitter or receiver coils, resulting in reflected power or changes in magnitude and phase \cite{buikman1988rf, Hess2018a, Vahle2020}. Load impedance is dominated by the impedance of the body at clinical field strengths \cite{buikman1988rf}. Additionally, motion can change tissue conductivity \cite{buikman1988rf} and distance between the coil and the body \cite{Navest2019}. RFC methods measure changes in load impedance passively or actively. A passive method known as a ``noise navigator" tracks changes in the variance of received noise \cite{Andreychenko2017}. Active methods measure reflected power at the transmitter \cite{buikman1988rf, Hess2018a} or voltage changes at the receiver coils due to the superposition of transmitted and induced EM fields (PT) \cite{Vahle2020}. In PT, the transmitter and receiver coils are distinct; thus, variations in distance between them can lead to additional magnitude and phase changes \cite{Speier_Bacher_2023}. 

In PT, a tone near the Larmor frequency is transmitted continuously and detected by the receiver coils. For some motion types, the transmitter and/or receiver coils must be close to the subject for high SNR sensing, such as cardiac motion \cite{Speier_Bacher_2023}. Due to its ability to sense many motion types with minimal hardware additions, PT has been integrated into commercial systems, offering valuable respiratory, cardiac, and head motion information \cite{Falcao2022, Vahle2020, Ludwig2020, Solomon2020, Wilkinson2021}. However, RFC methods are fundamentally limited in motion sensitivity, as measured by percent modulation: some studies suggest between 5 and 15\% modulation for breathing \cite{buikman1988rf, Navest2019} and 1-2\% for cardiac motion \cite{buikman1988rf, Speier_Bacher_2023}. Moreover, these methods may not translate well to low-field scanners. Because body impedance is weaker at lower frequencies, RFC methods may suffer from lower SNR in low-field systems \cite{buikman1988rf}.

In contrast, BPT is not tied to the Larmor frequency. Rather than transmitting a single RF tone, BPT uses two tones and exploits a nonlinear property of the receiver chain known as \textit{intermodulation} for reception. BPT can be separated into two pieces: the transmit field (BPT-Tx) and the received signal after mixing (BPT-Rx).
Two tones with frequencies that may lie far outside the MR BW are transmitted into the bore, creating two BPT-Tx fields. The BPT-Tx fields interact with the bore and body, are sensed by the receiver coil elements, and serendipitously mixed by intermodulation, creating a new signal called an intermodulation product or distortion (IMD) that we denote as BPT-Rx. Intermodulation is a property of nonlinear devices (e.g., diodes, transistors, etc.) in which inputs at frequencies $f_1$ and $f_2$ can create an output at a new frequency $f_{BPT}$ \cite{maas2003nonlinear}, where:

\begin{equation}
f_{BPT}= mf_2 + nf_1,
\end{equation}

\noindent and $m$ and $n$ are signed integers.
$f_1$ and $f_2$ are chosen such that the IMD at $f_{BPT}$ will fall close to the Larmor frequency. For example, two tones transmitted at $f_1 = 2.4$ GHz, $f_2=2.5278$ GHz will have an IMD at frequency
\begin{equation}
f_{BPT}= f_2-f_1 = 127.8 \text{ MHz},
\end{equation}

\noindent which is within our GE MR750W (GE Healthcare; Waukesha, WI) 3T scanner bandwidth. This example uses two tones to create two BPT-Tx fields, resulting in a single BPT-Rx via second-order intermodulation. Both BPT-Rx and PT appear as a peak in the Fourier transform of each \textit{k}-space line that is separable from the MR data (Figure \ref{fig:bpt_acq}a).

\subsection{Hardware Implementation}
\label{sec:hw}
\begin{figure}[!htb]%
\centering
\includegraphics[width=0.5\textwidth]{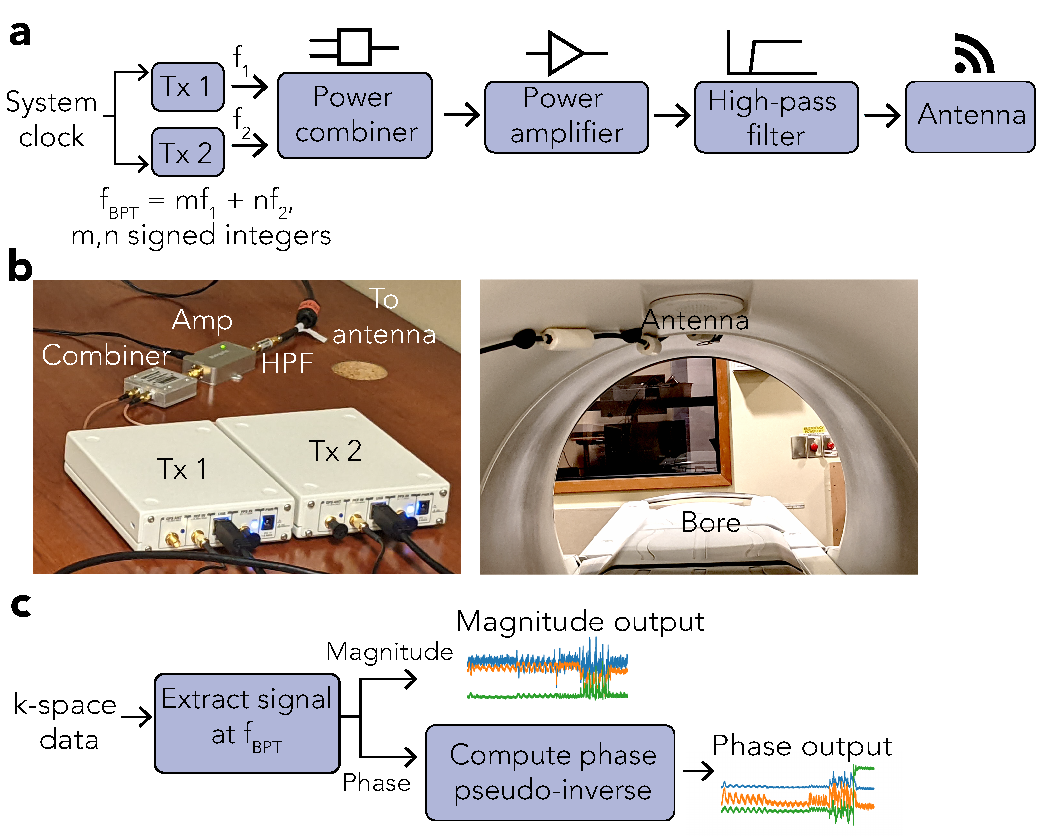}
\caption{\textbf{Acquisition and processing pipeline}. a) The acquisition pipeline, in which two tones were generated with two transmitters (e.g., software-defined radios (SDRs)), combined, amplified, high-pass filtered, and transmitted with an antenna. b) Photo of a sample setup, in which the transmitters (left) are USRP B200 SDRs and an antenna (right) is placed at the top of the bore. c) Basic reconstruction pipeline, where the complex signal at $f_{BPT}$ is extracted from the raw k-space data. The phase is computed by taking the pseudo-inverse of all possible reference coils relative to a chosen reference coil (Supporting Information).}
\label{fig:hw}
\end{figure}

BPT-Tx can be implemented with off-the-shelf and low-cost hardware (Figure  \ref{fig:hw}). We used two software-defined radios (SDRs) to produce the two tones (Ettus Research B200; National Instruments, TX, USA), synchronized to the system $10$ MHz clock. The tones were combined, amplified, filtered and transmitted using an antenna placed inside or outside the bore. The filter removed intermodulation from the transmit amplifier in the transmitted tones. The specific hardware components are listed in Tables S1 and S2.

\subsection{Intermodulation Properties}
\label{sec:intermod}
Intermodulation may occur by any non-linear component in the receiver chain. This could be passive, like diodes used for coil detuning and protection,  or  active, like the preamplifier circuits (preamps) themselves. We hypothesize that the preamps are the likely culprit because little BPT-Tx transmit power ($<20$dBm at the transmit antenna) is sufficient to induce IMD at a similar amplitude to the MR signal. We measured the strength of the IMD on a representative MR preamplifier (Clinical MR Solutions, LLC; WI, USA; Figure S1). Using a spectrum analyzer (FieldFox N9918A; Keysight Technologies; CA, USA), the input power was swept from $-14$dBm to $+2$dBm, and the power of the IMD was measured, along with the output power at $2.4$ and $2.5278$ GHz (“fundamental”). Since the MR signal power ranges from $-75$ to $-40$ dBm at $3$T \cite{sporrer2017fully}, Figure S1 shows that it is possible to obtain an IMD power close to that of the MR signal ($>-70$ dBm) with little BPT-Tx power ($-10$ dBm), suggesting that intermodulation in the preamp is the likely mechanism of BPT-Rx.



\subsection{EM Coil Motion Simulations} \label{sec:sim}
\begin{figure}[!htb]%
\centering
\includegraphics[width=0.6\textwidth]{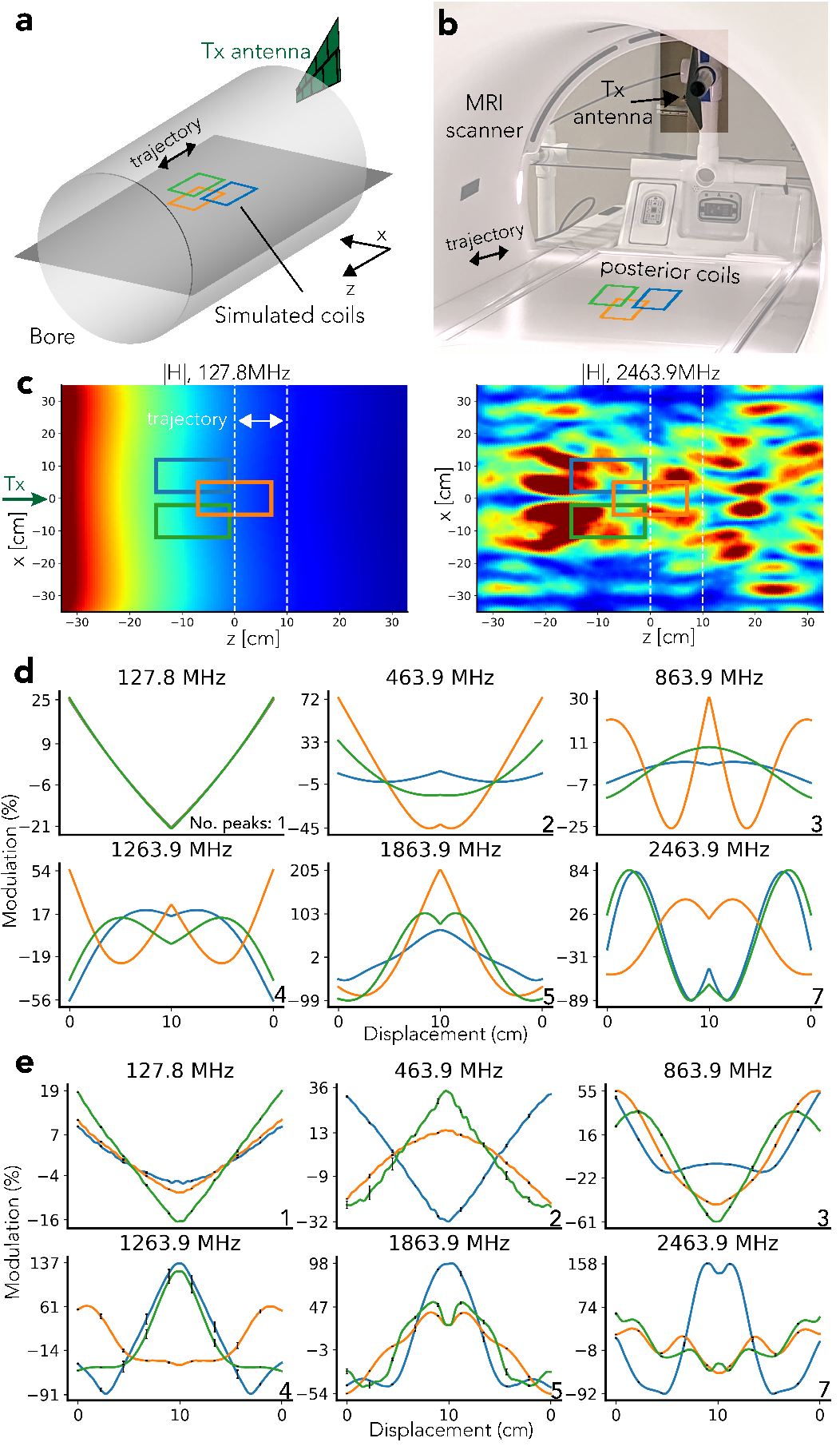}
\caption{\textbf{BPT finite-element EM field simulations}. a) Simulation setup for an idealized bore. Flux was computed through rectangular coils moving along the z-axis. b) The experimental setup. c) $|H|$ at the central slice of the bore at (left) $127.8$ MHz and (right) $2463.9$ MHz (product of $|H|$ at $2400$ and $2527.8$ MHz). Coil locations are shown by the colored rectangles, range of motion by white dashed lines and tx antenna location by the green arrow. While $|H|$ at $127.8$ MHz decays over distance, $|H|$ at $2463.9$ MHz shows standing wave patterns. d) Simulated magnitude of BPT-Rx in percent modulation units, with the maximum theoretical number of peaks on the bottom right.
e) The measured magnitude of BPT-Rx for three coils in the posterior array. The signals were averaged over 5 periods, with standard deviation in black. The number of peaks and level of modulation increase with frequency, showing the same trends in simulation and experiment. 
}
\label{fig:rocker}
\end{figure}
\clearpage

BPT-Rx, like PT and other RFC methods, senses changes in the EM field. However, when using high frequencies ($\sim 2.4$ GHz), it is additionally sensitive to EM wave effects. We simulated these effects in a basic simulation where BPT-Rx was computed across a range of frequencies in an ideal vacuum-filled bore, with three virtual coils moving along the z-axis. We validated the accuracy of the simulation in an experiment. Though the simulation and experiment do not capture the complexity of BPT on a human subject, they show the key differences between BPT and RFC methods: namely, that standing wave patterns are produced by the boundary conditions of the setup, these waves are sampled and mixed to create BPT-Rx, and that this mechanism may enable tunable sensitivity to motion depending on the transmit frequencies.

The simulations were conducted with the finite element solver High Frequency Structure Simulator (HFSS; Ansys, Canonsburg, PA, USA). The field was transmitted by an antenna outside the bore (Figure \ref{fig:rocker}a) and sensed by three virtual coils. The coils were moved $10$ cm back and forth along the z-direction (Figure \ref{fig:rocker}a). The bore was simulated as a perfect electrical conductor (PEC) and sized to match the GE 3T MR750W system used for all experiments ($70$ cm in diameter, $135$ cm in length). Its calculated cutoff frequency is $251.1$ MHz \cite{tang2011cutoff}. The simulated antenna was a wideband log periodic antenna ($600$ MHz --- $6$ GHz) modeled after the antenna used in experiments (Agatige; China). The EM field was simulated as a Multi-Frequency simulation with frequencies $127.8$, $400$, $527.8$, $800$, $927.8$, $1200$, $1327.8$, $1800$, $1927.8$, $2400$, and $2527.8$ MHz.

By the Maxwell-Faraday law of induction, the received voltage is proportional to the flux $\Phi = \int \vec{H} \cdot \vec{dA}$. Consequently, PT was simulated by computing $\Phi$ at $127.8$ MHz, and BPT-Rx by computing the product of $\Phi$ through the three coils for each BPT-Tx frequency pair. The coils were modeled as rectangular surfaces that were matched in size ($10 \times 14$ cm) and position to the posterior coils used in experiments (Figure \ref{fig:rocker}b). The magnitude of $\Phi$ was converted to percent modulation by dividing by the mean, i.e. for a signal $\mathbf{x}$ of length $N$,
\begin{equation}
    \%\mathrm{mod} = (\frac{\mathbf{x}}{\frac{1}{N}\sum{\mathbf{x}}} - 1)\times 100.
\end{equation}

In our experimental validation, we measured PT and BPT-Rx at the same frequencies and displacements as the simulated values. We used the hardware setup in Figure \ref{fig:hw}a to transmit BPT-Tx with a wideband amplifier (ZX60-83MP-S+; Minicircuits; NY, USA), and a $\sim$50dB Larmor frequency band-stop filter (ZX75BS-125-S+; Minicircuits; NY, USA). The BPT-Tx powers ranged from 5dBm to 10dBm at the antenna. To transmit PT, we used a waveform generator (SDG6022X; Siglent Technologies; OH, USA) and a power of -28dBm. We inductively supressed common-mode parasitics from the transmit cable using air-core winding of the coaxial transmission line. We moved a posterior receiver array back and forth along the z-axis using Rocker, an application that moves the scanner bed bidirectionally using the built-in motor (Figure \ref{fig:rocker}b). To ensure repeatability, we repeated the motion for 5 periods, aligned the signals, and plotted the mean signal, with black error bars indicating standard deviation. Magnitude BPT signals were extracted, low-pass filtered with a cutoff of $2$Hz, then converted to percent modulation.

Figure \ref{fig:rocker}c and d show the results of the simulation. The main differences between BPT and PT are due to the nonlinear reception of BPT-Rx and the frequency-dependent characteristics of the MRI bore, which acts as a cylindrical waveguide \cite{Brunner2009}. The simulated $|H|$ at $127.8 $MHz (Figure \ref{fig:rocker}c, left) decays with distance from the transmitter because it is below the waveguide cutoff frequency \cite{Brunner2009}. However, at frequencies greater than the cutoff, such as $2.4 $GHz, the transmitted tone produces a standing wave (BPT-Tx) that forms spatially-varying patterns (Figure \ref{fig:rocker}c, right) \cite{Brunner2009}. 

After evaluating the simulation qualitatively and quantitatively, we found that key features of the simulation are validated by experiment. In simulation and experiment,  the number of peaks, level of modulation, and spatial variation in PT and BPT-Rx signals increase with frequency (Figure \ref{fig:rocker}d and e). The first subplot is PT at $127.8$ MHz; the remaining plots are BPT-Rx, labeled by the average of the two transmit frequencies. The simulated PT transmit field decays in space and has no nonlinear sensing; consequently, there is a single peak in the received waveform, and adjacent coil signals appear similar in shape and level of modulation. This also holds true in the measured PT (Figure \ref{fig:rocker}e). However, in BPT-Rx, interaction with the bore and nonlinear sensing cause the predicted number of peaks to increase over frequency (explained in the Supporting Information). The maximum peak count (bottom right of each subplot in Figure \ref{fig:rocker}d and e)  is consistent with the simulated and experimental data. Moreover, because each coil samples the local field, adjacent coil signals appear to have different shapes and levels of modulation from one another. Finally, there is an increase in the level of modulation over frequency. At $127.8$MHz, the simulated modulation is within +/- $25\%$, while for $2463.9$MHz it ranges from $-89$ to $84\%$. This is a $3.7 \times$ increase in sensitivity. While the measured data shows a similar modulation range for PT, the measured BPT-Rx range at $2463.9$ MHz appears larger than the simulated range. Discrepancies are discussed in Section \ref{sec:discussion}. 

\subsection{Coil Vibration Experiment} \label{sec:sensitivity}
We performed a phantom experiment to evaluate the sensitivity of BPT-Rx to small motions by measuring vibrations of the receiver coil across a range of BPT-Tx transmit frequencies. Using the same transmit antenna and a combined anterior/posterior semi-flexible array (``GEM" anterior array; GE Healthcare; Waukesha, WI), we caused the coils to vibrate by placing the array on a plastic structure and moving it abruptly. The scanner cradle was moved at the maximum speed ($100$ mm/s) and a small displacement ($1$ cm), then stopped for 5 seconds for vibrations to decay.  PT and BPT-Rx were measured sequentially and continuously during both cradle motion and rest periods. The data was low-pass filtered with zero group delay and cutoffs of $1$ Hz for frequencies $<863.9$ MHz and $5$ Hz for frequencies $\geq 863.9$ MHz for better visualization. 

We validated the origin of these fluctuations by comparing BPT-Rx at the two highest transmit frequencies to displacement calculated from accelerometer measurements. A tri-axial SCL3300 accelerometer (Murata Electronics; Kyoto, Japan) was controlled by an Arduino Pro Mini (Arduino; Somerville, MA, USA) and synchronized to BPT and PT via a Transistor-Transistor Logic (TTL) signal from the scanner. The displacement $\Delta d$ was computed from acceleration by high-pass filtering with a cutoff of $2.5$ Hz, then integrating the signal twice. BPT-Rx data were low-pass filtered with a cutoff of $5$ Hz. Accelerometer and BPT-Rx measurements were obtained sequentially in order to prevent accelerometer noise from corrupting the BPT-Rx acquisition. RF excitation and gradients were turned off.

\subsection{Volunteer Experiments} \label{sec:volunteer}
To explore BPT's motion sensitivity in volunteers, we performed volunteer experiments with respiratory, bulk, cardiac, and head motion. The coil arrays used in these experiments varied in rigidity and level of contact with the subject. While the coils in the respiratory, bulk, and cardiac experiments touched and moved with the subject's body, the coils in the head motion experiment were fixed and not in close contact with the subject, potentially leading to different mechanisms of motion sensing. The BPT transmit antenna(s) were placed away from the subject at the top and/or side of the bore, while many of the sensors used for comparison (PT antenna, PPG, ECG, and accelerometer) were placed on the subject's chest or abdomen. All scans were acquired on healthy volunteers after obtaining IRB approval and informed consent. Floating cable traps were placed on the cable connected to the transmit antenna(s) to suppress common-mode currents induced by the MRI transmit RF field \cite{seeber2004floating}. The received levels of BPT-Rx and PT were equalized by adjusting the transmit powers until the root-sum-square of the amplitude of BPT-Rx and PT peaks across coils were approximately equal. The received levels for different BPT-Tx transmit frequencies were equalized in the same way, resulting in a range of transmit powers. All filtering had zero group delay.

The maximum BPT-Tx power used (100 $mW$) would produce energy absorption of 0.08 $mW/cm^2$ in the worst case, which is well below the limit of 5 $mW/cm^2$ given by the Federal Communications Commission \cite{fcc2020}. We estimated this by assuming isotropic radiation from the antenna at a distance of 10 cm from the subject, as in the head motion experiment. The power density $P_d$ is then 100 $mW$ divided by the surface area of a sphere with radius $r=10$ cm:
\begin{equation}
    P_d = 100 / (4 \pi r^2) = 0.08 mW/cm^2
\end{equation}

\subsubsection{Respiratory Motion Experiment}
A volunteer was asked to perform different breathing types: chest breathing, stomach breathing, rapid-shallow breathing, and breathing with simultaneous bulk motion of the chest (Figure \ref{fig:resp}). The bulk motion consisted of rotation of within +/- 3 degrees about the head-foot axis and translation between -15mm to 5mm in the left-right and anterior-posterior axes (Figure S2). PT and BPT-Tx were transmitted simultaneously at frequencies of $127.8$ MHz and $2.4/2.5276$ GHz, respectively. The scan was acquired with the semi-flexible GEM anterior array. The data was low-pass filtered and converted to percent modulation units (excluding coils with low means, for which the percent modulation is artificially large). The phase was extracted using the method described in the Supporting Information. The low-pass filter cutoff was $2$ Hz for respiratory motion (Figure \ref{fig:resp}a) and $14$ Hz for bulk motion (Figure \ref{fig:resp}c) to better visualize sharp movements.

We further investigated separation between motion types by applying Principal Component Analysis (PCA) to PT and BPT-Rx. We compared the time evolution of three PCs of PT and BPT-Rx for a portion of the data with breathing and bulk motion (Figure \ref{fig:resp}e). Three PCs were chosen for visualization purposes.

We repeated this experiment with the same volunteer at BPT-Tx frequencies $300/427.6$, $800/927.6$, $1200/1327.6$, and $1800/1927.6$ MHz, with simultaneous PT at $127.8$ MHz throughout. BPT-Tx transmit powers ranged from 10 dBm to 20 dBm at the antenna, while PT was fixed at -36 dBm at the antenna in order to result in the same received level (explained in Section \ref{sec:volunteer}). Results from selected coils are shown in Figure S3.

\subsubsection{GEM Cardiac Motion Experiments}
To evaluate frequency-dependent cardiac sensing with BPT, we acquired a series of breath-held cardiac scans of a volunteer at BPT-Tx frequencies ranging from $400$ MHz to $2.4$ GHz (Figure \ref{fig:cardiac}), chosen to be the same as the EM simulation (Figure \ref{fig:rocker}). The scans were acquired on the same GEM anterior array coil used in the respiratory experiment but with 16 additional posterior coils (GE Healthcare; Waukesha, WI). BPT-Tx and PT were transmitted simultaneously, with PT at 127.6 MHz and -38 dBm and BPT-Tx ranging from 10 to 17 dBm. The scans were acquired with RF excitation and gradients off in order to avoid artifacts (see Section \ref{sec:eddy}). BPT-Tx was transmitted using a 2.4 GHz printed circuit-board (PCB) dipole antenna placed at the top of the bore, while PT was transmitted with a waveform generator (SDG6022X; Siglent Technologies; OH, USA) and a small broadband loop placed on the chest. We used a wideband amplifier and band-stop filter for the BPT-Tx setup (Section \ref{sec:sim}). BPT-Rx and PT were acquired with simultaneous photoplethysmogram (PPG) on the subject's finger as a reference for signal timing and shape. 

The PT/BPT magnitude signals were low-pass filtered with a cutoff of $25$ Hz, mean-subtracted, and normalized to unit variance (Figure \ref{fig:cardiac}a). Signals from four coils with the greatest energy in the cardiac frequency range [0.9, 3] Hz were plotted. We further compared the signal characteristics of BPT-Rx and PT (Figure \ref{fig:cardiac}b). PT from an empirically chosen coil element was low-pass filtered with a cutoff of 2 Hz, while BPT-Rx was filtered with a cutoff of 25 Hz for best visualization.

We performed an additional cardiac experiment to assess whether rigid contact with the coil was necessary to sense BPT-Rx signal at 2.4 GHz. We measured breath-held cardiac BPT-Rx data while the GEM coil was not in physical contact with the subject. A 2.4 GHz dipole PCB antenna at the top of the bore transmitted the BPT at 20 dBm and 2.4/2.5278 GHz. The results of this experiment are in Figure S4.

\subsubsection{dBCG Validation}
We compared BPT-Rx from breath-held scans of a volunteer to dBCG acquired with a tri-axial accelerometer placed on the chest as well as to PT, ECG, and PPG \cite{Anand2023} (Figure \ref{fig:dBCG}). The accelerometer setup (Figure  \ref{fig:dBCG}b) was the same as in the coil vibration experiment, but with the accelerometer on the subject's chest. We used a 2D SPGR sequence with RF and gradient excitation turned off, BW=$62.5$ kHz 
and TR=$8.7$ ms to allow data transfer from the Arduino. As in the cardiac motion experiment, a semi-flexible GEM receiver coil was used, and PPG was measured from the subject's finger. ECG was also acquired with electrodes on the chest. The accelerometer signals were high-pass filtered with a cutoff of $4$Hz, then integrated twice to obtained displacement.

We performed a least-squares fit to compare dBCG and BPT-Rx quantitatively by regressing the multi-coil BPT-Rx signals to dBCG measured on the left-right axis. Magnitude BPT data were low-pass filtered with a cutoff of $15$Hz, and all signals were de-meaned before the regression. We denote the resulting signal as BPT-dBCG (Figure \ref{fig:dBCG}e). 

\subsubsection{AIR Coil Cardiac Motion Experiment}
To investigate the effect of coil characteristics on BPT-Rx, we additionally measured cardiac signal from a breath-held volunteer scan using a flexible coil array marketed as the AIR coil (GE Healthcare; Waukesha, WI). BPT-Tx and PT were transmitted simultaneously at frequencies 1.8/1.9298GHz and 129.8MHz, with transmit powers of approximately 10dBm and -44 dBm at the antenna, respectively. 129.8 MHz was chosen because PT modulation is empirically more pronounced than at 127.8MHz yet still in the receiver BW. BPT-Tx was transmitted using the setup described in Figure \ref{fig:hw} and a 4G LTE antenna, while PT was transmitted with a waveform generator (SDG6022X; Siglent Technologies; OH, USA) and a small broadband loop, with both BPT and PT transmitters placed on the coil. Data was acquired with RF and gradient excitation turned off. Results from a single coil are shown in Figure S5.

\subsubsection{Head Motion Experiment}
We measured head motion on a volunteer with simultaneous MIMO BPT and PT while acquiring a series of low-resolution 3-dimensional images on a rigid 22-channel head-neck coil (GEM HNU; GE Healthcare; Waukesha, WI) that was not touching the subject. The arrangement of coil elements is shown in Figure S7. The goals of the experiment were two-fold: first, to evaluate the sensitivity and accuracy of BPT-Rx and PT in detecting head motion, and second, to introduce and evaluate MIMO capability for both BPT-Rx and PT.

For the first goal, we explored whether PT and BPT-Rx could capture the lower-dimensionality of nodding and shaking out of the larger space of possible head positions. We performed a calibration scan and an inference scan. In the calibration scan, the subject moved in a raster fashion to sample all translations and rotations within +/- 3mm and +/- 4 degrees (Figure S6a). During the inference scan, the subject shook their head ``no" and nodded their head ``yes" (Figure S6b). Both scans were acquired using a 3D SPGR sequence (FOV=42 $\times$ 42 $\times$ 27cm, FA=5, BW=250kHz, acceleration=2.5$\times$). The calibration scan was acquired at $\sim1$ frame per second with 115 frames, and the inference scan at $\sim0.4$ frames per second with 50 frames.

The transmit setup consisted of two 2.4 GHz dipole PCB transmit antennas placed on the top and side of the head coil with cardboard in between the antenna and the coil to prevent detuning (Figure \ref{fig:head}a). The frequencies were 2500/2372.39, 2400/2527.78 for BPT-Tx and 127.58, 127.81 MHz for PT with transmit powers of +15dBm and -42 dBm respectively.  Two SDRs and two dual-channel RF transmit devices (SynthHD; Windfreak Technologies, LLC; FL, USA) were used to transmit all tones. The SDRs and one SynthHD produced two sets of BPT-Tx, while the second SynthHD produced the PTs. Each BPT-Tx chain consisted of a power combiner, narrow-band WiFi amplifier, and high-pass filter (Figure \ref{fig:hw}). An additional combiner (ZFSC-2-372-S+; Minicircuits; NY, USA) was used to combine the two BPT-Tx tones with a PT, yielding three tones on each antenna.

To evaluate the accuracy of BPT-Rx and PT, we obtained ground-truth motion estimates by registering the low-resolution image timeseries. The registration estimates were Savitzky-Golay filtered (window=11 samples, order=4) and compressed to 3 PCs for visualization purposes. The PCs (reg-PCs) explained $99 \%$ of the variance of the data. An affine registration, recording translation and rotation, was performed using the SimpleElastix Python package with default settings \cite{marstal2016simpleelastix}.

The calibration BPT-Rx and PT data were first preprocessed by averaging over each frame and filtering using a Savitzky-Golay filter (window=11, order=4) to reduce noise. Each data set had a dimension of $[115\times 22 \times 2]$ corresponding to number of frames, coils and tx antennas, respectively. It was then reshaped to $[115\times 44]$ and compressed to 3 PCs (Figure S6c-d). The inference BPT-Rx and PT data were processed in the same way and then projected onto the learned PCs. The resulting signals explained $\sim$95\% of the data variance for both BPT and PT. As in Figure \ref{fig:resp}e, we plotted each time point in the PC-space (Figure \ref{fig:head}d). 

To evaluate the MIMO setup quantitatively, we regressed the multicoil BPT-Rx and PT signals to each reg-PC using data from both antennas, top alone, and side alone. A single reg-PC was chosen for plotting (Figure S8), while the correlations with all reg-PCs are reported in Table S3 (bolded values indicate the maximum values). Qualitative plots of the time and PC domain plots using data from both antennas, top alone, and side alone are shown in Figure S9.

\subsection{SNR}
We performed an image SNR comparison to ensure that BPT-Rx does not adversely impact image SNR. A uniform phantom was scanned with BPT-Tx, PT, and no BPT-Tx/PT using a 2D SPGR sequence (TR=$34$ms, FA=$30$, resolution=$0.9$mm, BW=$250$kHz, FOV=$50$cm) and a panel antenna placed at the top of the bore. The frequencies and powers were 2.4/2.5278 GHz and 17 dBm for BPT-Tx and 127.8MHz and -40dBm for PT. BPT-Rx and PT were cropped out of the image. An SNR map was computed using the method developed by Kellman et al \cite{kellman2005image}. 

\section*{Results}
\subsection{Coil Vibration Experiment}
\begin{figure*}[htb]%
\centering
\includegraphics[width=0.6\textwidth]{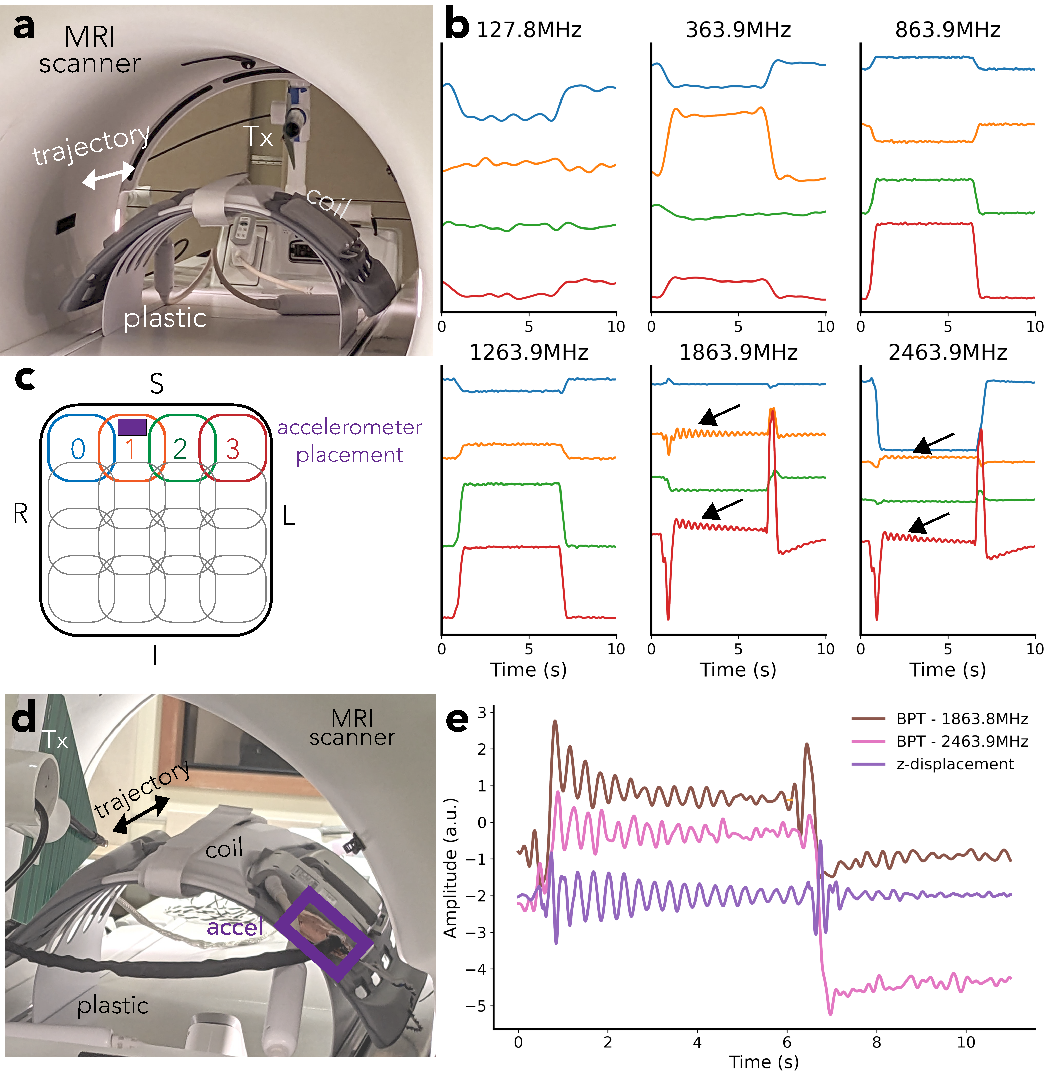}
\caption{\textbf{Coil vibration measurement and validation.} a) The cradle was moved at maximum speed and stopped for 5 seconds. b) The first period of the PT (top left) and BPT-Rx (remaining plots) is displayed after low-pass filtering. At frequencies $>1.2$ GHz, BPT-Rx senses the vibration of the coils due to the abrupt motion of the cradle (black arrows). c) The arrangement of the coils. d) The setup of the accelerometer experiment, with the accelerometer (accel) indicated by the purple box. e) The measured BPT-Rx signal at the two highest frequencies vs. displacement ($\Delta$d) measured sequentially with accel placed on coil 1. The signals appear to match.}
\label{fig:vibration}
\end{figure*}
\clearpage

Figure \ref{fig:vibration}b shows the results of the coil vibration experiment, with setup in Figure \ref{fig:vibration}a and physical coil arrangement in Figure \ref{fig:vibration}c. At the two highest frequencies ($1.839$ GHz and $2.4639$ GHz), the coil signals show small fluctuations (black arrows in Figure \ref{fig:vibration}b). We hypothesized that these are vibrations of the coil when the cradle suddenly moves.

We validated this hypothesis by comparing BPT-Rx at the two highest frequencies to displacement measured by the accelerometer. Figure \ref{fig:vibration}e suggests that the displacement (purple) is very similar to BPT-Rx at 1.8 (brown) and 2.4 GHz (pink) in shape and vibration frequency. Moreover, after the cradle is moved in the opposite direction (6-10 seconds), the vibration appears to be dampened compared to the earlier movement (1-6 seconds) in both the accelerometer measurement and BPT-Rx. Therefore, it appears that BPT-Rx at high frequencies is able to capture small vibrations of this semi-flexible receiver coil. 
\subsection{Volunteer Experiments}

\begin{figure*}[!htb]%
\centering
\includegraphics[width=0.6\textwidth]{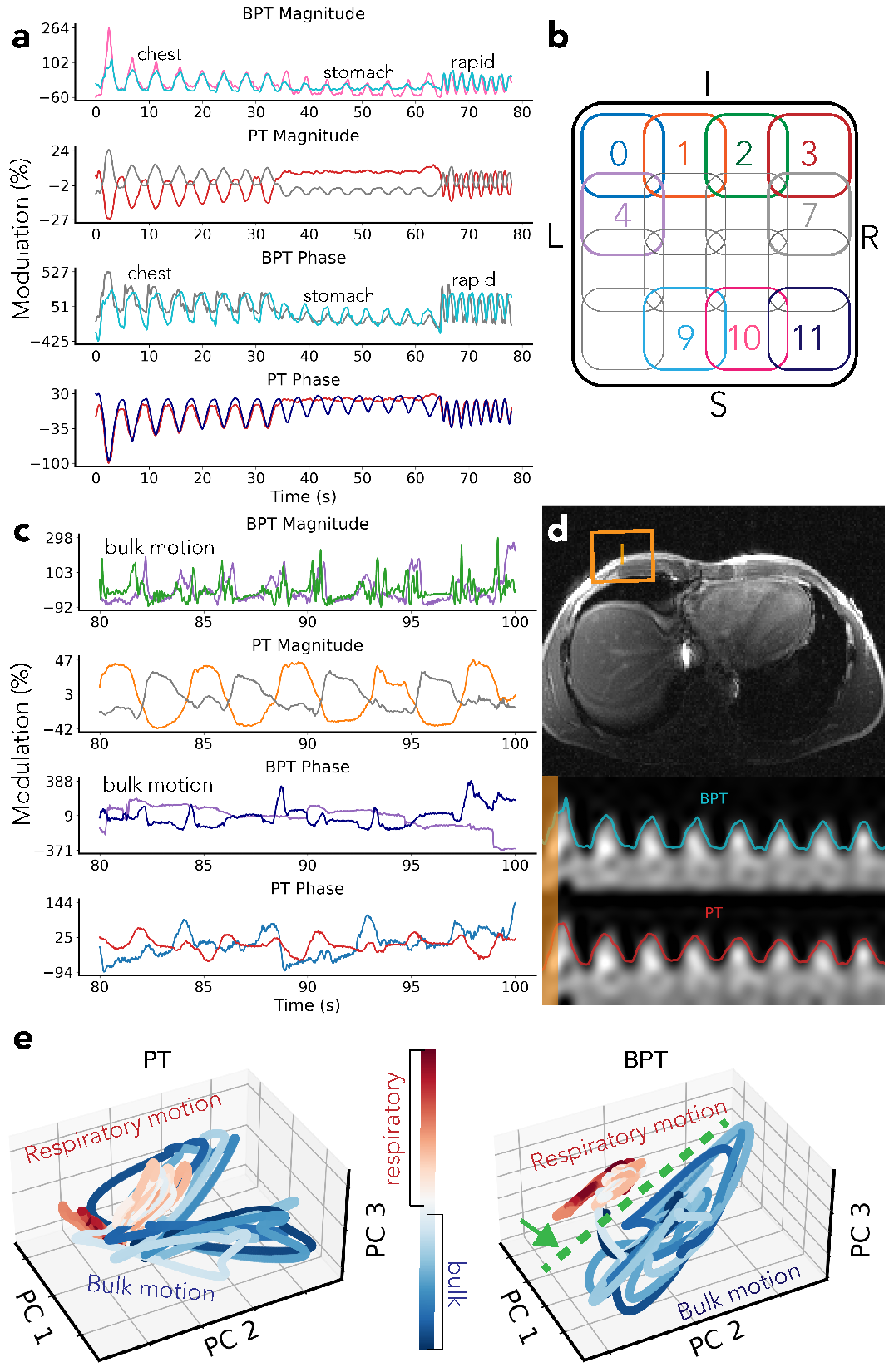}
\caption{\textbf{Respiratory motion sensing with BPT and PT}. a) BPT-Rx and PT signals during chest, stomach, and rapid breathing, chosen from the two most modulated coils and displayed in percent modulation units. BPT-Rx modulation is 6 to 7 times greater than PT. b) Coil arrangement and colors corresponding to a), c), and d). c) BPT-Rx magnitude and phase modulation (first and third row) and PT magnitude and phase modulation (second and last row) during bulk motion. BPT-Rx modulation is larger and sharper than the PT. d) BPT-Rx (top) and PT (bottom) magnitude overlaid on a patch of the image (orange box) for coils 9 and 3, respectively. Both appear to match the displacement of the patch qualitatively. e) The time evolution of the three main PCs of the PT and BPT-Rx during breathing and bulk motion. The green arrow and line indicate a separating plane between breathing and bulk motion. 
}
\label{fig:resp}
\end{figure*}
\clearpage

\subsubsection{Respiratory Motion Experiment}
The respiratory motion experiment suggests that BPT-Rx is sensitive to changes in breathing and bulk motion (Figure \ref{fig:resp}). Figure \ref{fig:resp}a displays the two most modulated coils from BPT-Rx and PT, each scaled within their maximum and minimum percent modulation. The numbered coils are physically arranged as in Figure \ref{fig:resp}b. We compared the range (max - min) of BPT-Rx and PT modulations. BPT-Rx is 6.3$\times$ more modulated than PT in magnitude and 7.3$\times$ in phase during breathing. When overlaid on a patch of the image over time, both BPT-Rx (top) and PT (bottom) breathing signals appear comparable (Figure \ref{fig:resp}d).

BPT-Rx is also 4.3$\times$ more modulated than the PT in the bulk motion portion of the experiment in magnitude and 3.2$\times$ in phase (Figure \ref{fig:resp}c). While the BPT-Rx magnitude shows much sharper peaks for bulk motion than breathing, the PT magnitude is smoother, similar to the breathing signal. Thus, Figure \ref{fig:resp}a and c suggest that BPT-Rx can achieve greater motion sensitivity than PT, and that bulk motion appears noticeably different from respiration.

Figure \ref{fig:resp}e further demonstrates that breathing is better separated from bulk motion in BPT-Rx PCs (right) compared to PT PCs (left), with the green dashed line indicating a separating plane. Separation in the PC feature space may be a good metric to distinguish between motion types. For instance, after a training period, this information could be used to reject outliers or prospectively acquire data only during respiratory motion.

We found that BPT-Rx signals from some coils over a broad BPT-Tx frequency range have multiple peaks. The simulation (Figure \ref{fig:rocker}) suggested that standing wave patterns with multiple peaks and nulls are created when transmitting into an empty bore. In a setting with a human subject in the bore, the body modulates the standing wave based on its conductive properties, and this standing wave changes with motion; thus, the setting is much more complex. Figure  S3a shows manually chosen coil signals from the respiratory experiment with multiple peaks across a broad range of BPT-Tx frequencies (300/427.6, 800/927.6, 1800/1927.6, 2400/2527.6MHz), with corresponding coil positions in S3b. This result further supports the standing wave mechanism.

\subsubsection{GEM Cardiac Motion Experiments}
\begin{figure*}[!htb]%
\centering
\includegraphics[width=0.8\textwidth]{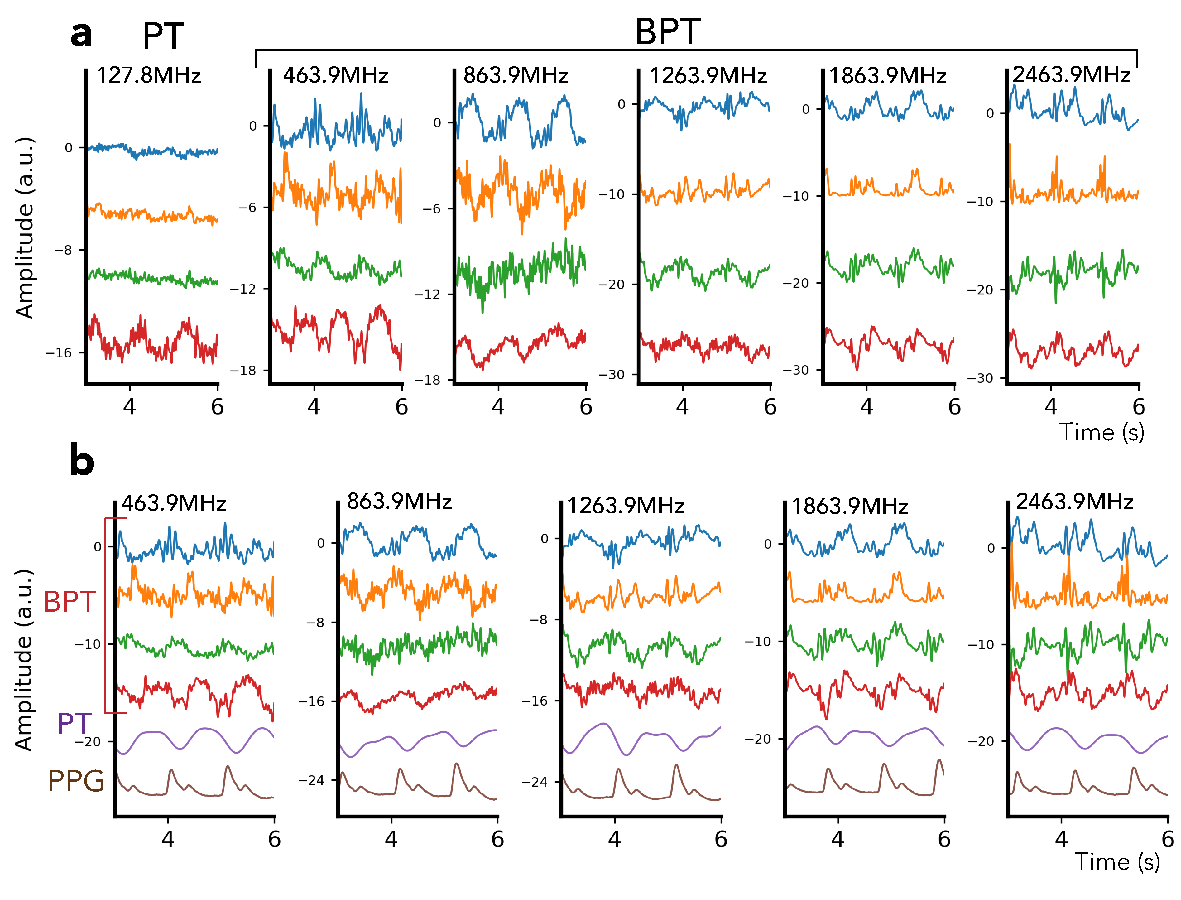}
\caption{\textbf{Cardiac motion sensing with PT and BPT.} Cardiac PT and BPT-Rx signals were obtained simultaneously from a healthy volunteer during breath-held scans across BPT-Tx frequencies. a) PT and BPT-Rx signals from the four coils with the greatest energy in the cardiac frequency range after low-pass filtering with a cutoff of 25 Hz, mean-subtraction, and normalization to unit variance. b) Low-pass filtered PT and BPT-Rx signals, along with simultaneous PPG (bottom). The filter cutoffs were 2 Hz for PT and 25 Hz for BPT.  The cardiac PT/BPT-Rx signal appears smooth at frequencies $< 1.2$ GHz but contains sharp peaks for higher frequencies, suggesting that BPT-Rx may be more sensitive to blood volume changes at lower frequencies and surface vibrations at higher frequencies.}
\label{fig:cardiac}
\end{figure*}
\clearpage

On the GEM semi-flexible coil array, BPT-Rx at low frequencies appears to reflect blood volume changes, while BPT-Rx at high frequencies appears more sensitive to surface vibrations of the body.
Figure \ref{fig:cardiac}a shows PT and BPT-Rx magnitudes from the four coils with the greatest energy in the cardiac frequency range. PT cardiac signal appears noisier and contains a single peak for each heartbeat, a signal shape that arises from blood volume changes \cite{Speier_Bacher_2023}. However, the BPT-Rx signals at high frequencies (1.8 and 2.4GHz) are cleaner and appear to contain multiple peaks. Figure \ref{fig:cardiac}b further shows the differences between PT, BPT-Rx, and PPG acquired simultaneously. At lower frequencies ($< 1.2$GHz), BPT-Rx changes smoothly. At higher frequencies ($\geq 1.2$GHz), BPT-Rx contains multiple sharp peaks. It appears that the sensing mechanism is different --- as explained in the next section, we hypothesize that BPT-Rx at high frequencies captures dBCG, which measures millimeter-scale vibrations of the body due to the ballistic forces of blood. Given that intermediate frequencies (464 and 864MHz) may contain more than one peak per cycle but have a similar shape to PT, it is possible that the cardiac signal arises both from both blood volume changes and dBCG. The improved quality of the signal at high frequencies suggests that BPT-Rx cardiac signal can be extracted from the raw data with minimal processing, which makes BPT a promising candidate for real-time cardiac gating applications.

However, Figure S4 suggests that rigid contact with the subject may be required to obtain these BPT-Rx results at high frequencies. When the coil is elevated above the subject (Figure S4b), the sharp peaks do not appear as visible (Figure S4d). There is still some cardiac modulation, but it appears weaker. We hypothesize the reasons for this in Section \ref{sec:air}.

\subsubsection{dBCG Validation}
\begin{figure*}[!htb]%
\centering
\includegraphics[width=0.5\textwidth]{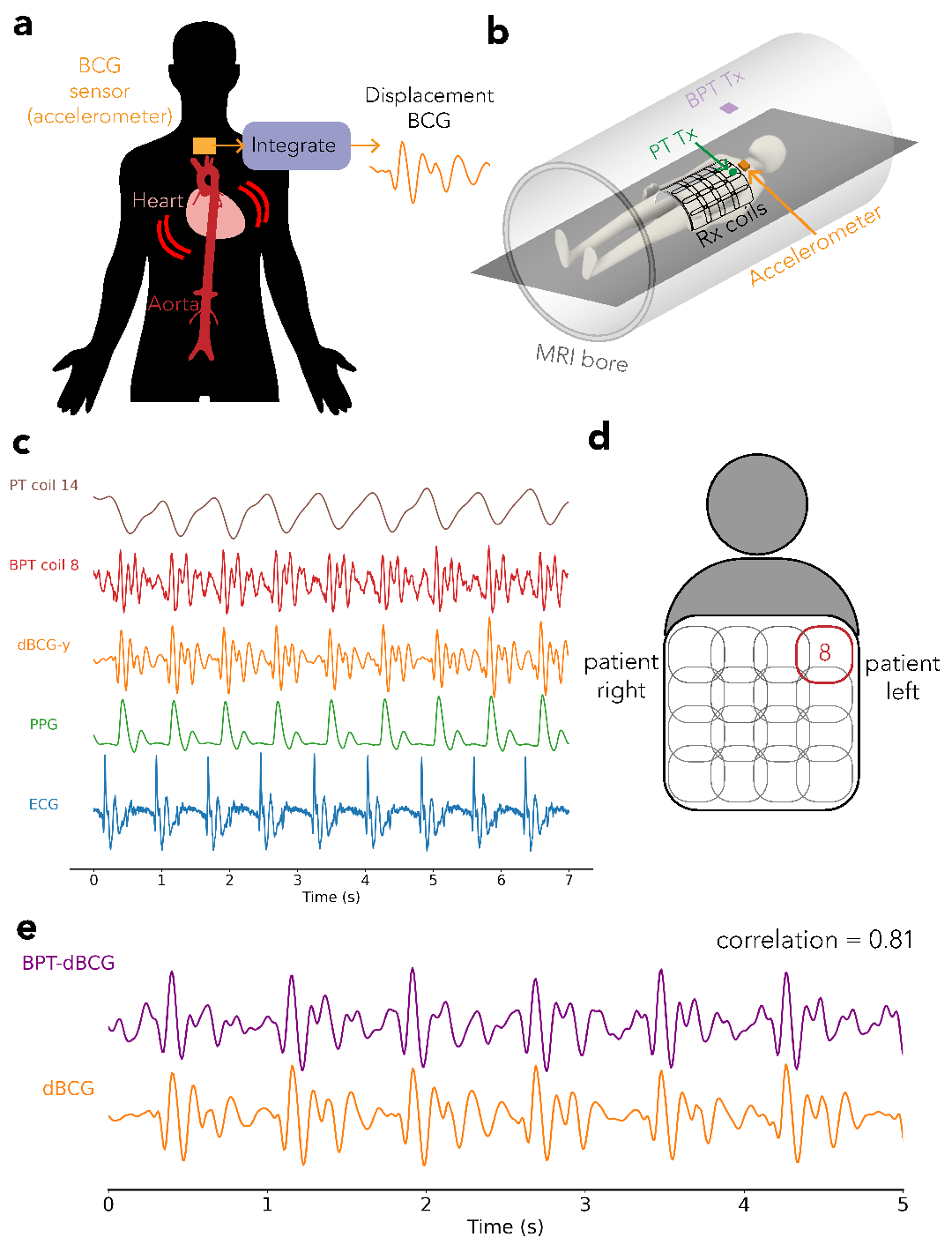}
\caption{\textbf{dBCG measurement and validation with BPT.} a) Displacement BCG (dBCG) measures the recoil of the body due to the ballistic forces of blood through the aorta. It was measured with a tri-axial accelerometer, then integrated twice to obtain displacement. b) Placement of the BPT tx antenna, PT tx antenna, and the accelerometer in the bore \cite{Anand2023}. c) Low-pass filtered PT (cutoff = 3 Hz), BPT-Rx signal from a single coil vs. computed dBCG, PPG, and ECG. The first wave of BPT-Rx and dBCG appear later than ECG but earlier than PPG. d) Physical location of the BPT coil corresponding to c). e) BPT signals were regressed to dBCG, then low-pass filtered; this is denoted as BPT-dBCG. BPT-dBCG correlates strongly with dBCG (correlation = $0.81$).}
\label{fig:dBCG}
\end{figure*}

Cardiac signals acquired by BPT-Rx at 2.4/2.5278GHz highly correlate with dBCG \cite{Brablik2022, noordergraaf1961physical, Anand2023} (Figure \ref{fig:dBCG}). This suggests not only that BPT-Rx at microwave transmit frequencies is sensitive to very small vibrations but also that BPT-Rx could offer information about the cardiac cycle that is complementary to MR imaging. dBCG can be measured in many ways, including fiber-optic sensors \cite{Nedoma2020}, pneumatic sensors \cite{Brablik2022}, cameras \cite{Shao2017}, doppler radar \cite{Pinheiro2010}, or accelerometers \cite{Shao2017}.

Figure \ref{fig:dBCG}c demonstrates the timing of the raw BPT-Rx signal from the coil in Figure \ref{fig:dBCG}d relative to low-pass filtered PT (cutoff = 3 Hz), dBCG, PPG, and ECG signals. The timing and peaks of the BPT-Rx signal qualitatively match dBCG. Quantitatively, the regressed BPT-dBCG has a strong correlation of 0.81 with dBCG (Figure \ref{fig:dBCG}e) \cite{Anand2023}. We hypothesize that BPT senses mechanical vibrations; therefore, BPT-dBCG may be enhanced due to the stiffness of the receiver array and contact with the subject.

\subsubsection{AIR Coil Cardiac Experiment} \label{sec:air}
One caveat, however, is that dBCG may not be visible with BPT-Rx on all coils. The AIR coil experimental results are shown in Figure  S5. The top subplot shows percent modulation of BPT-Rx and PT relative to the mean. While BPT-Rx is approximately twice as modulated as PT, the modulation is still within +/- 1\%, which is much smaller than the level of modulation in dBCG ($\approx 30 \%$ \cite{Anand2023}). Moreover, the signal shape and timing appear very similar between PT and BPT-Rx. We hypothesize that BPT-dBCG is visible in the experiments with the GEM coil because of its mechanical stiffness, which amplifies vibrations of the body when touching it. In contrast, the AIR coil is flexible and may dampen these vibrations. Figure S4 further supports this hypothesis - when the GEM coil is not in physical contact with the subject and thus not vibrating, BCG does not appear visible.

The cardiac results suggest that BPT may be influenced by blood volume changes as well as surface vibrations of the body and the coil. The relative contribution of each of these mechanisms appears to depend on the stiffness of coil, the level of contact with the subject, and the transmit frequencies.

\begin{figure*}[!htb]%
\centering
\includegraphics[width=0.7\textwidth]{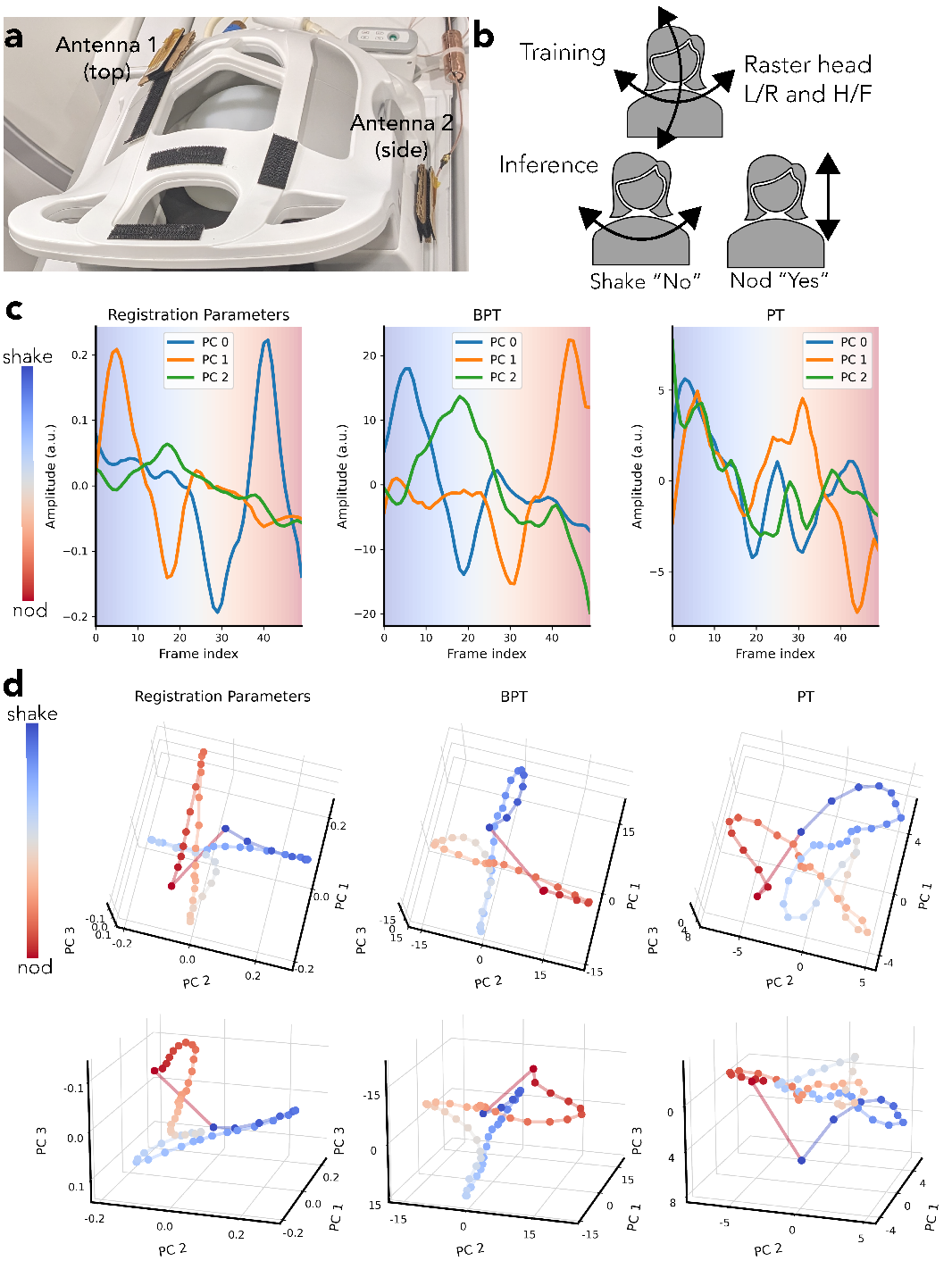}
\caption{\textbf{MIMO head motion sensing with BPT}. a) Two antennas were placed at the top and side of the head coil, each transmitting at two BPT-Tx tones and one PT. b) During a calibration or training scan, the volunteer rotated and translated their head in a raster fashion. During an inference scan, they shook “no” (left-right) and nodded “yes” (up-down). c) The three main principal components (PCs) of the ground-truth registration parameters (reg-PCs), BPT-Rx and PT plotted on the same scale versus time and d) in the PC feature space viewed from two different angles (top, bottom). The two head motions (shake “no”, nod “yes”) show the same “x” shape in param and BPT-Rx, with the crossing point corresponding to the common head position between nodding and shaking. However, PT does not show the same structure or smoothness.}
\label{fig:head}
\end{figure*}
\clearpage

\subsubsection{Head Motion Experiment}

Figure \ref{fig:head}c shows the three PCs for the registration parameters (reg-PCs), BPT-PCs, and PT-PCs over time, while Figure \ref{fig:head}d shows the same data in the PC space from two viewing angles.
The smoothness of the BPT-PC curve in Figure \ref{fig:head}c suggests that the BPT-PCs may be less noisy than PT. In \ref{fig:head}d, the BPT-PCs overall look very similar to the reg-PCs. Both BPT-PCs and reg-PCs show a crossing behavior that corresponds to the position of the head between nodding and shaking. Moreover, both show smooth trajectories for nodding and shaking. However, the PT-PCs are noisier, with less clear structures to distinguish nodding from shaking.

Quantitatively, MIMO appears to have benefits for BPT-Rx and PT: the regressed BPT-Rx and PT (Figure S8, Table S3) computed with data from both antennas have greater correlation with the reg-PCs than regressed BPT-Rx/PT using data from a single antenna. Regressed BPT-Rx (Figure S8b) has consistently higher correlation than PT (Figure S8a) over all reg-PCs (Table S3). Figure S9 shows the time and PC-domain plots for the top and side antenna data alone, showing qualitative differences between the two datasets. These results suggest that MIMO-BPT has the potential for accurate quantitative rigid motion correction.

\subsection{SNR} \label{snr}
\begin{figure}[!htb]%
\centering
\includegraphics[width=0.6\textwidth]{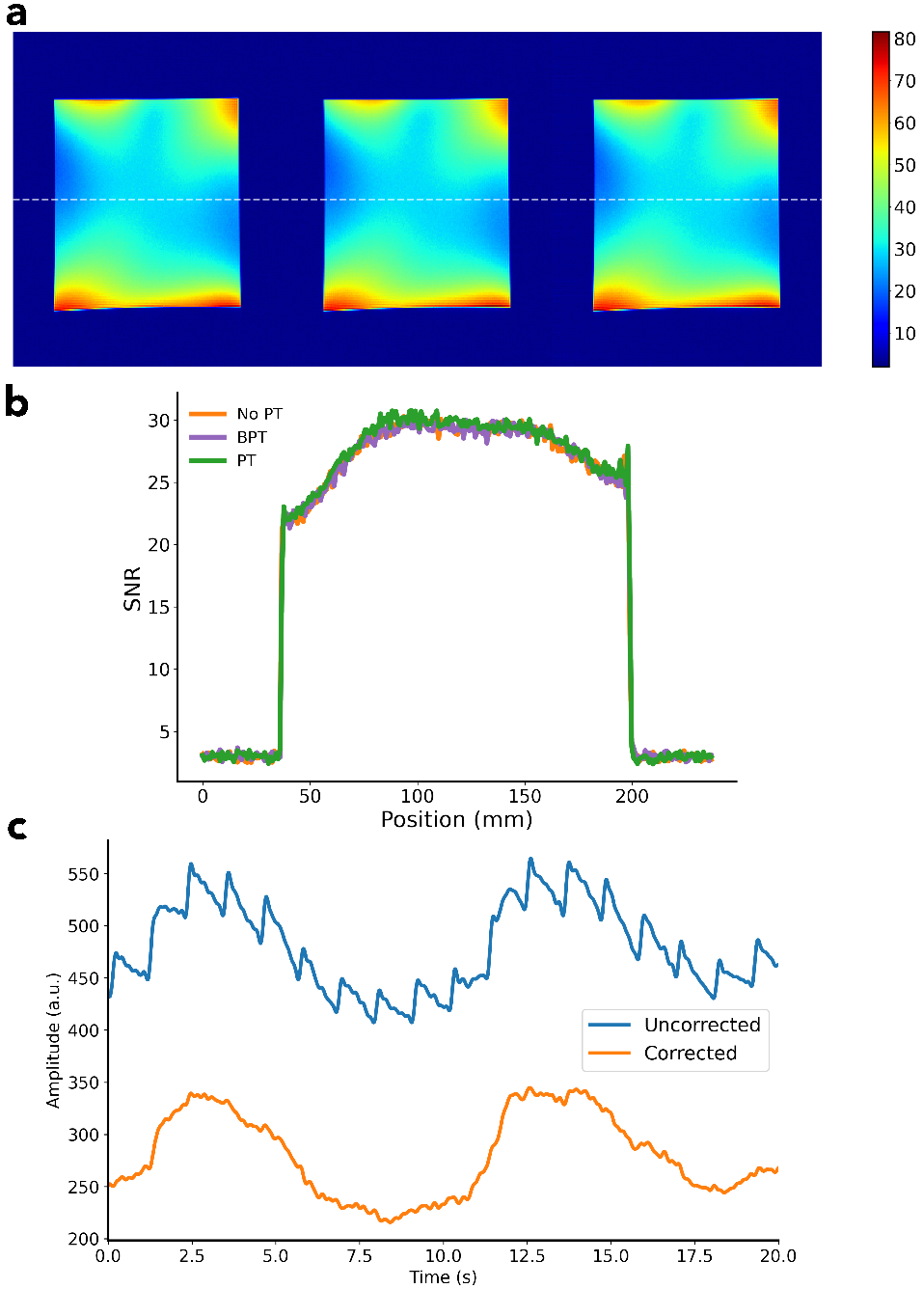}
\caption{\textbf{SNR and Artifacts}. a) Image SNR maps were computed in three acquisitions with no PT, BPT, and PT using a uniform phantom with a 32-channel body coil. b) Line plots through the center of the maps, indicated by the dashed line. c) Switching gradient fields can cause the antenna to vibrate, resulting in artifacts in BPT-Rx (top, blue), which was measured with BPT-Tx frequencies of 2.4/2.5478 GHz. These can be corrected (bottom, orange) by a linear fit, as described in the Supporting Information.} 
\label{fig:snr}
\end{figure}

The SNR maps (Figure \ref{fig:snr}a) and line profiles (Figure \ref{fig:snr}b) are nearly identical with and without BPT or PT. This suggests that PT and BPT-Rx have no adverse impact on SNR. For the transmit power levels used in the volunteer experiments and SNR maps, there is no visible receiver gain compression.

\subsection{Vibration Artifacts} \label{sec:eddy}
The high sensitivity of the BPT also causes it to be sensitive to other vibrations of the system (Figure \ref{fig:snr}c). We hypothesize that the antenna and receiver coils are caused to vibrate by mechanical and electrical coupling to the MRI scanner. Mechanically, structures in the scanner vibrate due to Lorentz forces on the coil windings from the static magnetic field ($B_0$). If the antenna or receiver array is placed on any vibrating surface, it will also vibrate. Electrically, the antenna and coils may experience Lorentz forces due to induced eddy currents from the switching gradient fields, causing them to vibrate. The electrical coupling occurs only if the antenna and coils are placed inside the MRI bore. This artifact can be mitigated by filtering, placing the antenna outside the bore, or dampening its vibration. It may also be approximated by using a linear fit per-coil and subtracting the estimate from the data, which is described in the Supporting Information.

\section*{Discussion} \label{sec:discussion}
In this paper, we present BPT, a simple system exploitation that enables simultaneous RF motion sensing with the MRI data acquisition. BPT has the potential to operate at any frequency and in any MRI scanner independent of field strength. Moreover, the transmitter setup is simple to implement. BPT uses RF standing waves to obtain percent modulation that can be more than 5 times greater than PT. By removing the requirement that the transmitted RF be tied to the Larmor frequency, BPT generalizes RFC methods and opens up new possibilities for frequency-dependent motion detection. Our preliminary results show that BPT can separate respiratory from bulk motion, capture dBCG at microwave frequencies on a semi-rigid coil, and operate as a MIMO system. 

Simulations suggest that BPT samples standing wave patterns in the bore (Figure \ref{fig:rocker}). These simulations agree well with experimental results. Differences between simulation and experiment could be due to antenna characteristics and additional structures in the bore, such as the cradle, cables, and additional coils in the bed, which are not modeled in the simulation for simplicity. There may also have been unsuppressed common-mode parasitics, which could have radiated from the transmit cable and caused additional reflections. Accurate modeling of the human body may be necessary to simulate the behavior of BPT on a human subject.

The simulation considers only the magnitude of the flux, $|\Phi|$. Though the voltage is the time derivative of the flux, because the fields are complex exponentials in time, the voltage is directly proportional to the flux, i.e. $V(t) = e^{j\omega t} \Phi(t)$. The scaling does not affect the magnitude of the signal. Therefore, we omitted it. We chose to compare only $|
\Phi|$ between simulation and experiment because there are possible contributions to the phase that were not part of the simulation, such as AM-PM modulation \cite{maas2003nonlinear}. Due to these effects, the level of modulation in the measured phase is greater than expected in all experiments.

Volunteer experiments suggests that BPT's sensitivity to motion depends on the transmit frequencies, antenna placements, and coil characteristics. Cardiac BPT-Rx on the semi-flexible GEM coil correlates strongly with dBCG acquired simultaneously using an accelerometer when the coil is moving with the subject (Figure \ref{fig:dBCG}). However, dBCG does not appear visible in BPT-Rx on the flexible AIR coil (Figure S5), nor on the GEM coil when it is not in contact with the subject (Figure S5). BPT-Rx on the AIR coil appears similar to PT in signal shape, timing, and level of modulation. The head motion experiment used a coil that was not in rigid contact with the subject, yet still showed high correlation with ground-truth motion estimates from image registration (Table S3). The contributions to BPT-Rx are thus complex and require further exploration to characterize fully.

We have demonstrated that BPT can be scaled to multiple antennas (Figure \ref{fig:head}), and MIMO may allow the ability to separate and learn different motion trajectories. MIMO-BPT appears more similar to the reg-PC curves, and achieves high correlation ($>0.97$) with the reg-PCs when regressed; however, further analysis is required to investigate whether it provides benefits for quantitative head motion correction compared to a single BPT-Tx or PT.

In all experiments, we compared BPT-Rx to PT and common peripheral sensors (ECG and PPG). We could have additionally compared BPT-Rx to other RFC methods such as reflected power from a parallel transmit array \cite{Hess2018a} or the noise navigator \cite{Andreychenko2017}. However, our MRI system does not have a parallel transmit array available. Moreover, the noise navigator is passive and thus has inherently lower SNR compared to active methods such as PT \cite{Navest2019}. Therefore, we determined that the most reasonable comparison would be to PT.

BPT exploits a nonlinear property of the preamps; however, this nonlinearity may differ between coils and systems, and thus be a potential barrier to general implementation. Some array coil electronics may have inductors in the receive path before the preamp. Then, frequencies at the GHz range will be attenuated, and sensing BPT-Rx may not be possible without altering the coils. We also chose to focus on second order intermodulation (i.e., $f_{BPT}= f_2-f_1$) in this work. A higher order may allow for a larger bandwidth of intermodulation; however, this may require greater transmit power to overcome attenuation by the receiver chain.

\section*{Conclusion}
BPT offers a rich dataset of motion information. Obtaining dBCG measurements with BPT could characterize cardiac function simultaneously to the MRI exam \cite{MARCH1955,blacher1999aortic}. BPT has the ability to distinguish between motion types, offering the potential for improved motion correction. Preliminary work has demonstrated the utility of BPT for retrospective head motion correction \cite{Huttinga2023}. Using BPT with the MRI system has many other potential applications, including cardiac and respiratory gating \cite{Lamar2022}; motion sensing for low-field systems, in which body impedance changes are minimal \cite{Chen2023, marques2019low}; sensing motion deep in the body such as fetal motion \cite{roy2019fetal}; or using multiple transmitters to reconstruct microwave images \cite{alon2021stroke}. For example, researchers in Shanghai Jiao Tong University have implemented BPT on a low-field $0.25$T MRI system to sense and correct for respiratory motion \cite{Chen2023}. The simplicity of implementation could enable widespread adoption in many MRI systems. 

\section*{Acknowledgments}
 The authors thank the ISMRM Reproducible Research Study Group for conducting a code review of the code (commit hash e7d11e9) supplied in the Data Availability Statement. The scope of the code review covered only the code’s ease of download, quality of documentation, and ability to run, but did not consider scientific accuracy or code efficiency. 
 
 We thank Katie Lamar-Bruno for her feedback on the writing and help with various experimental setups, including the log-periodic antenna apparatus and the cabling for the accelerometer validation. We thank Julian Maravilla for providing the coil model used in the simulation in Figure  1 and the head coil model renders in Figure S7. We additionally thank Efrat Shimron for editing the paper and Alan Dong for interesting discussions. We acknowledge Karthik Gopalan, Jason Stockmann, and Jonathan Polimeni for their code to compute image SNR. We acknowledge support from GE Healthcare, the NSF GRFP fellowship, and NIH grants U01EB025162, R01HL136965, and U01EB029427.

\subsection*{Author contributions}
S.A. and M.L. conceptualized the work. S.A. carried out the electromagnetic simulations, experiments, and data analysis. Both authors discussed the results and contributed to the manuscript.

\subsection*{Financial disclosure}
None reported.

\subsection*{Data availability}
The datasets generated and analyzed during the current study are available via \url{https://doi.org/10.5281/zenodo.10967226}. The code is publicly available on \url{https://github.com/mikgroup/bpt_paper}, with the latest commit hash as 0e80ed0.

\subsection*{Conflict of interest}
M.L. receives research support from GE Healthcare. M.L is a founder, board member and stock owner of InkSpace Imaging, which could benefit from this study.


\section*{Supporting information}
The following supporting information is available as part of the online article:

\noindent
\textbf{BPT Phase Processing.}

\noindent
\textbf{Standing Wave Calculations.}

\noindent
\textbf{Vibration Artifact Removal.}

\noindent
\textbf{Table S1.}
{Type and placement of BPT and PT antennas: this table shows the type and placement of antennas for BPT and PT. Double dashes ('--') indicate that the antenna or placement was the same, e.g., the same log periodic antenna was used to transmit both BPT and PT.}

\noindent
\textbf{Table S2.}
{$2.4$GHz BPT hardware.}

\noindent
\textbf{Table S3.}
{PC correlation values. We computed the Pearson correlation coefficient of each registration parameter PC (reg-PC) with each BPT/PT PC computed from both antennas, the top antenna, or the side antenna and report the max correlation value (for instance, reg-PC 0 has a maximum correlation with BPT-PC 0 of 0.45 computed from both antennas).}

\noindent
\textbf{Figure  S1.}
{\textbf{Intermodulation measurements on a preamp}. The strength of the intermodulation product (IMD) was measured for two tones at 2.4GHz (left) and 2.5278GHz (right) on a preamplifier interface box for custom receiver coils (Clinical MR Solutions, LLC; WI, USA). Using a spectrum analyzer (FieldFox N9918A; Keysight Technologies; CA, USA), the input power was swept from -14dBm to +2dBm, and the power of the IMD was measured, along with the output power at 2.4 and 2.5278GHz (“fundamental”). The second-order intercept point (IP2) where the lines cross was extrapolated based on fitting lines to the fundamental and IMD data. The measurements suggest that it is possible to obtain an IMD power close to that of the MR signal ($>-70$dBm) with little BPT-Tx power ($-10$dBm) - suggesting that intermodulation in the preamp is the likely mechanism of BPT-Rx, and that there may not be significant gain suppression so far from the IP2 point.}

\noindent
\textbf{Figure  S2.}
{\textbf{Bulk motion in the respiratory experiment}. Bulk motion estimates from the respiratory experiment, with rotation angle on the left, and displacements on the right. The estimates were obtained by registering the images using SimpleElastix \cite{marstal2016simpleelastix}, a rigid transformation, and default rigid registration options (mutual information as the metric; multi-resolution registration; stochastic gradient descent). The rotation was within +/- 3 degrees, and the displacement was between -15 and 5 mm.}

\noindent
\textbf{Figure  S3.}
{\textbf{BPT-Rx respiratory signal over frequency}. A volunteer performed different breathing types (chest breathing, stomach breathing, rapid-shallow breathing, and breathing with simultaneous bulk motion of the chest). The experiment was repeated at BPT-Tx frequencies of 300/427.6MHz, 800/927.6MHz, 1200/1327.6MHz, 1800/1927.6MHz, and 2400/2527.6MHz. a) BPT-Rx signals with multiple peaks were manually chosen from each experiment, with coil positions in b). There appears to be at least one coil signal with multiple peaks for each of the frequencies, except for 1200/1327.6MHz. These peaks could be due to the shape of the BPT-Tx standing wave patterns.}

\noindent
\textbf{Figure  S4.}
{\textbf{Rigidly contacting vs elevated GEM coil cardiac experiment}. – breath-held cardiac BPT-Rx signals were acquired on a volunteer with a) the GEM anterior array (AA) coil placed on the chest and b) elevated from the chest using a plastic support. BPT signals from selected AA coils in c) the normal configuration and d) the elevated configuration. The elevated signals show lower cardiac modulation, and perhaps no longer a dBCG signal.
}

\noindent
\textbf{Figure  S5.}
{\textbf{AIR coil results}. BPT and PT were acquired simultaneously in a breath-held scan with RF excitation and gradients off using an abdominal AIR coil. The BPT frequencies were 1.8/1.9298GHz, and PT was 129.6MHz. Top: BPT and PT are plotted for a single coil in percent modulation units after being low-pass filtered with a cutoff of $5$Hz. Bottom: The same BPT and PT data after band-pass filtering between $0.5$ and $5$Hz. Both appear comparable in terms of signal shape and modulation level.}

\noindent
\textbf{Figure S6.}
{\textbf{Motion parameters during the MIMO head motion experiment}. In the calibration scan, the subject attempted to capture the full rigid parameter space (i.e., all possible translations and rotations within an approximate range of +/- 3mm and +/- 4 degrees) by moving in a raster fashion. In the inference scan, they shook their head “no” and nodded “yes”. PT and BPT data were combined across antennas and averaged over frames. Rigid motion parameters estimated from registering the images from a) the calibration scan and b) the inference scan. PT (top) and BPT (bottom) from the four most modulated coils are plotted from the c) the calibration scan and d) the inference scan.}

\noindent
\textbf{Figure  S7.}
\textbf{GEM head coil arrangement}. a) rendering of the GEM Head and Neck Unit (HNU) coil used for the head motion experiment. b) Approximate physical arrangement of the 22 coil elements. Elements 0, 1, and 2 belong to the posterior array integrated into the scanner bed.

\noindent
\textbf{Figure  S8.}
\textbf{BPT and PT regression to reg-PC 1.} Filtered and concatenated data were regressed to reg-PC 1 for a) PT and b) BPT-Rx data. The top row is for data from the top antenna, the second row for the side antenna, and the bottom row for both. Pearson correlation coefficient is reported for each subplot. The correlation is higher for PT and BPT-Rx when using data from both antennas and is higher for BPT-Rx than for PT.

\noindent
\textbf{Figure  S9.}
{\textbf{PCs with different antennas vs combined.} a) BPT and PT PCs computed only from the antenna at the top. The PCs from registration were unchanged. b) BPT and PT PCs from the top antenna in the PC space. c) BPT and PT PCs computed from the side antenna vs time and d) in the PC space.}


\vspace*{6pt}
\appendix

\clearpage

\listoffigures

\end{document}